\documentstyle[editedvolume,psfig,namedreferences]{crckapb}

\def\la{Ly$-\alpha~$}
\def\simleq{\widetilde <}
\def\kms{km\ s$^{-1}$}
\def\etal. {{\it et. al.}}

\begin{opening}
\title{A radio Search for high redshift HI absorption}
\subtitle{}

\author{Jayaram N. Chengalur, A. G. de Bruyn, R. Braun}
\institute{NFRA, P.O. Box 2, 7990 AA Dwingeloo, The Netherlands}
\author{C. L. Carilli}
\institute{Center for Astrophysics, Cambridge, MA, USA}

\end{opening}

\runningtitle{High z HI absorption}

\begin{document}


\section{Introduction}
	The rare damped-\la systems seen in absorption against the UV
continuum of high redshift QSOs nonetheless dominate the observed neutral
gas at high  redshift. HI 21cm absorption  studies of these systems are
challenging and only a handful of have been detected hitherto
(Carilli 1994). HI 21cm observations are however of great value because
they potentially yield  physical information (velocity dispersion, spin
temperature, size), that cannot  be obtained by other means.
\par
	If some damped-\la systems contain a significant amount of dust,
then QSOs behind such systems will be considerably extincted. Further,
since high resolution optical spectroscopy is only practical for bright
QSOs, such extincted QSOs will be excluded from observing samples, leading
to an underestimate of the number count of damped-\la systems, and hence
also of the gas content of the early universe. If the dust obscuration
is sufficient, the number counts of QSOs themselves could be strongly
biased, and if the damped-\la population is sharply divided into two
halves, one with little or no dust and another with substantial dust,
(if for example systems above a certain redshift are much dustier than
those at lower redshifts) this bias could in fact be completely
undetectable in purely optical studies (Heisler \& Ostriker 1988).
Such a bias however would be apparent when trying to find optical
counterparts to radio sources, and the fact that optical identification
was possible for a complete sample of radio sources (Shaver, 1994) makes
this unlikely.
\par
	Although the effect of obscuration by dust may not be as substantial
as envisaged by Heisler \& Ostriker, it could nonetheless be important. Two
independent observations in fact provide complementary evidence for the
presence of dust in high redshift damped-\la systems, (i)~QSOs with
damped-\la systems along the line of sight have redder UV continuum than
those without, (Pei \etal. 1991)~(ii)~The gas phase Cr  abundance in these
systems is much lower than that of Zn, (Steidel \etal. 1995). In the galaxy
Cr is depleted onto dust grains while Zn is not. Thus the count of both
damped-\la systems and that of QSOs is biased, however since the spectral
properties of the dust are poorly constrained, the severity of the bias
is correspondingly uncertain. Fall \& Pei (1995) estimate that up to 70\%
of high redshift quasars could be missed.
\par
	Blind radio searches (against the lobes of radio galaxies  which
are often well removed from the central core, however note that since one
requires to detect the \la line to determine the redshift, the line of sight
to the central object itself must be relatively dust free), are less subject
to the bias introduced by dust extinction and are hence of particular value
in resolving this controversy. Such searches have traditionally been
difficult because of the challenging instrumental requirements and the
hostile radio interference environment at low frequencies. In this paper
we describe a pilot radio search for damped-\la systems made using a novel
observing mode at the WSRT and present preliminary results.

\section{Observations and Data Reduction}
\subsection{Observations}
	An absorber with column density N$_{\rm H} > 2\times 10^{20}$ cm$^{-2}$
is encountered on the average 0.25 times along a line of sight
a unit redshift long and centered at $z \sim 2.5$ (Lanzetta \etal. 1991).
For a column density N$_{\rm H} > 1.0\times 10^{21}$ the corresponding
probability is~0.09. Note that this probability estimate includes only those
absorbers seen against optically bright QSOs and is hence a lower
limit if dusty systems do indeed exist. Further, the estimate is
based largely on QSOs with redshift~$<~3$, and there is some
evidence that the probability of encountering a high column
density absorption system increases (by a factor of 2 - 3)
at a redshift of $\simeq$ 3. (White, Kinney \& Becker 1993). However,
even in the most optimistic scenario, the number count is low enough
to make it essential to search  a large redshift path interval before
the probability of detecting a damped-\la system becomes meaningful.
The velocity width of the HI 21cm absorption signal is probably small,
$10 - 100$ \kms (however, any object which large enough to cover the
entire radio emitting region is unlikely to have very small velocity
width).  This combination of high spectral resolution and
simultaneously high instantaneous bandwidth is quite challenging to
achieve. Further the interference environment at these frequencies
($200 -300$~MHz) is often hostile. The new Compound Interferometry
(CI) observing mode at the WSRT, along with the newly commissioned
broadband 92cm system (Carilli \etal. 1995) however does make such
observations feasible.

	In CI mode, the WSRT is split into two phased arrays, and the
summed signal from these arrays is cross-correlated, i.e. one has a
two-element interferometer, with each element being a phased array.
The reduction in the number of measured spatial baselines (from 40
to 1) allows one to achieve high spectral resolution, up to 8192
channels across an instantaneous bandwidth of 20~MHz. In practice
this 20~MHz bandwidth is obtained by using 4 contiguous 5~MHz bands,
and after allowing for overlap between the bands the usable instantaneous
bandwidths is 16.4 MHz. There is reduced sensitivity to interference
because of the interferometers rejection of terrestrial signals, and
further unlike single dish radio spectroscopy there is no need to spend
large amounts of time calibrating the total power induced spectral band
pass shape. A first round of observations were conducted in March~1995,
when a total redshift interval of 3.5 was observed towards 4 objects.
The typical integration time per frequency setting was $\sim 8000$ seconds.
Software limitations prevented us from attaining the highest possible
resolution; we were instead restricted to a resolution a factor of
two worse (i.e. $\sim 25$ \kms).

\subsection{Data Reduction}
	The shape of the spectral baseline (or the equivalently the
frequency dependence of the visibility) is a function of the distribution
of background sources and the hour angle. (For example, a bright source at
the 10~dB point of the primary beam would lead to baseline structure on
the scales of $\sim 2$ MHz). Figure~\ref{fig:visshape}[A] shows
the observed visibility during a single 80s integration towards the
radio source 8C1435, showing the dramatic influence of background sources
in determining the shape of spectrum. However, this spectrum can be easily
modeled (Figure~\ref{fig:visshape}[B]) if one has a map of the sky (as
seen by the same telescopes). Modeling is done by special purpose software
produced by us, and is in general quite successful
although there are occasionally residuals which might be attributable to
imperfect knowledge of the shape of the primary beam, and also perhaps
to some low level cross-talk in the adding stage.

	The data reduction proceeds along the following steps (i) the raw
spectra are calibrated to an absolute flux level using observations of
calibrator sources interspersed throughout the observation (ii) the spectra
are then corrected for the instrumental bandpass using observations of these
same calibrator sources (iii) Model background sources  (obtained from an
independent continuum map of the field, sometimes from the WENSS survey,
and sometimes from other projects) are subtracted from the spectrum,
(iv) any residual large scale baseline features are removed by low order
polynomial fitting (v) RFI is flagged and the spectra are co-added to yield
the final spectrum.

\section{ Preliminary Results}
	Figure~\ref{fig:8c143} shows the final spectrum towards 8C1435,
which is a radio galaxy at $z\sim 4.25$ (Lacy etal. 1994). It has a
flux at 350~MHz of $\sim 2.7$ Jy and a spectral index $\alpha \sim -1.2$.
The low frequency flux is presumably dominated by the two hot spots,
which have a separation of $\sim 5^{''}$, or $\sim$ 20~kpc in a flat
$\Omega=1$ universe. The total redshift range observed is $\Delta z \sim
0.63$, which is about 65\% of the available redshift (using the WSRT
broadband 92cm system) towards this object. The noise is $\sim 13$~mJy,
which is the expected thermal noise limit. The spectral resolution is
$\sim 25$ \kms.

	We do not detect any narrow linewidth absorption in this redshift
interval. At the center of the band  the $4\sigma$ upper limit to the optical
depth (assuming that the object covers the entire radio emitting region)
is $\tau \simleq 2\times10^{-2}$, or
\begin{equation}
	N_H \simleq 8.8\times 10^{20}\times {\Delta V \over 25
	{~\rm km s^{-1}}} \times {{\rm T}_s \over 1000 {~\rm K}}
\end{equation}

	The data to the remaining objects is being reduced. A further
redshift interval of $\sim 3.5$ will be observed shortly (with a frequency
resolution a factor of two better than that for these observations).
Continuum mapping observations of the same field
with the broadband  92cm should enable us to  model the spectra
within the thermal noise and thus obviate the necessity to fit low order
polynomials and dramatically improve our ability to detect both weak broad
lines, (for example from larger scale structure along the line of sight),
and also recombination lines from ionized gas associated with the radio
galaxy itself.
\par
	The WSRT will soon have a UHF(high) system  covering the
frequency range between 1200~MHz and 700~MHz, with a system temperature
$\sim 75$~K at the upper end of the band, which will make it practical
to search for HI 21 cm absorption towards complete samples of radio
galaxies. Predictions that the maximum effect due to obscuration is
at $z \sim 1$, (Fall \& Pei 1995) makes such a search specially
interesting. {\bf Acknowledgments.} These observations would not have
been possible without the substantial and enthusiastic support of the
WSRT staff, in  particular A. J. Boonstra, A. Bos, J. Bregman, H. Butcher,
H. v. Sommeren Greve and the telescope operators. We are also grateful
for software and insightful comments from F. Briggs.

\section{References}
\def\refindent{\advance\leftskip by 24pt \parindent=-24pt}
\def\journal#1#2#3#4#5{{\refindent
                      {#1}        
                      {#2},       
                      {#3\/},     
                      {#4},       
                      {#5}        
                      \par }}
\def\circular#1#2#3#4{{\refindent
                     {#1}          
                     {#2},         
                     {#3\/},       
                     {#4}          
                     \par}}
\refindent{Carilli C. L.}{1994}{J. Astrophy. \& Astron.}
\circular{Carilli C. L. \etal. }{1995}{The 92cm broadband system at the WSRT}
	{Technical Note}
\circular{Fall M. S. \& Pei, Y. C.}{1995}{To Appear in QSO Absorption Lines}
	{Symposium Proceedings}
\journal{Heisler J. \& Ostriker J. P.}{1988}{Ap.J.}{332}{543}
\journal{Lacy, M. et. al.}{1994}{M.N.R.A.S.}{271}{540}
\journal{Lanzetta K. M., Wolfe A. M., Turnshek D. A., Lu L.,
	    McMahon R. G. \& Hazard C.}{1991}{Ap.J.S.}{77}{1}
\journal{Pei Y. C., Fall, S. M., \& Bechtold J.}{1991}{Ap. J.}{378}{6}
\circular{Shaver P. A.}{1995}{17th Texas Symposium on Relativistic
	Astrophysics \& Cosmology} {Symposium Proceedings}
\journal{Steidel C. C., Bowen D. V., Blades J. C. \& Dickinson M}
        {1995}{Ap. J.}{440}{L45}
\journal{ White R. L., Kinney, A. L. \& Becker R. H.}{1993}{Ap.J.}{407}{456}

\begin{figure}
\caption{}
\label{fig:visshape}
\end{figure}

\begin{figure}
\caption{}
\label{fig:8c143}
\end{figure}

\end{document}